\documentclass[10pt,letterpaper]{article}

\usepackage[top=0.85in,left=2.75in,footskip=0.75in]{geometry}
\usepackage{amsmath,amssymb}
\usepackage{changepage}
\usepackage[utf8x]{inputenc}
\usepackage{textcomp,marvosym}
\usepackage{cite}
\usepackage{nameref,hyperref}
\usepackage[right]{lineno}
\usepackage{microtype}
\DisableLigatures[f]{encoding = *, family = * }
\usepackage[table]{xcolor}
\usepackage{array}

\usepackage[T1]{fontenc}
\usepackage{lmodern}

\usepackage{amsmath,amssymb}
\usepackage{xcolor}
\usepackage{siunitx}

\newcolumntype{+}{!{\vrule width 2pt}}

\newlength\savedwidth

\raggedright
\usepackage[aboveskip=1pt,labelfont=bf,labelsep=period,justification=raggedright,singlelinecheck=off]{caption}

\makeatletter
\renewcommand{\@biblabel}[1]{\quad#1.}
\makeatother

\usepackage{lastpage,fancyhdr,graphicx}
\usepackage{epstopdf}
\fancyheadoffset[L]{2.25in}
\fancyfootoffset[L]{2.25in}
\newcommand{\fieldBPF}{E_{\mathrm{BFP}}}
\newcommand{\R}{\mathbb{R}}
\newcommand{\I}{\mathcal{I}}

\begin{document}
\vspace*{0.2in}
\begin{flushleft}
{\Large
\textbf\newline{Localization of fixed dipoles at high precision by accounting for sample drift during illumination}
}
\newline
\\
Fabian Hinterer\textsuperscript{1,\Yinyang},
Magdalena C Schneider\textsuperscript{2,3,\Yinyang},
Simon Hubmer\textsuperscript{4},
Montserrat López-Martínez\textsuperscript{2},
Ronny Ramlau\textsuperscript{1,4},
Gerhard J Schütz\textsuperscript{2}
\\
\bigskip
\textbf{1} Johannes Kepler University Linz, Institute of Industrial Mathematics,
Linz, Austria\\
\textbf{2} Institute of Applied Physics, TU Wien, Vienna, Austria\\
\textbf{3} Janelia Research Campus, Howard Hughes Medical Institute, Ashburn, VA, USA\\
\textbf{4} Johann Radon Institute Linz, 
Linz, Austria\\

\bigskip
\Yinyang These authors contributed equally to this work.

* fabian.hinterer@indmath.uni-linz.ac.at (FH), schneider@iap.tuwien.ac.at (MCS) 
\end{flushleft}

\section*{Abstract}
Single molecule localization microscopy relies on the precise quantification of the position of single dye emitters in a sample. This precision is improved by the number of photons that can be detected from each molecule. It is therefore recommendable to increase illumination times for the recording process. Particularly recording at cryogenic temperatures dramatically reduces photobleaching and thereby allows a massive increase in illumination times to several seconds. As a downside, microscope instabilities may well introduce jitter during such long illuminations, deteriorating the localization precision. In this paper, we theoretically demonstrate that a parallel recording of fiducial marker beads together with a novel fitting approach accounting for the full drift trajectory allows for largely eliminating drift effects for drift magnitudes of several hundred nanometers per frame. 
\smallskip

\noindent \textbf{Keywords.} Single molecule localization microscopy, cryogenic temperatures, fluorescence microscopy, sample drift, drift correction

\section{Introduction}

Single molecule localization microscopy (SMLM) techniques, including STORM \cite{Rust_2006,Huang_2008} and PALM \cite{Hess_2007}, rely on the temporal separation of the fluorescent emission from dye molecules, resulting in the sequential imaging of only a sparse subset of emitters in each frame. The position of the emitters can then be localized with arbitrary high precision, limited mainly by the signal-to-noise ratio (SNR) of the data. Recording SMLM data in cryogenic conditions improves the SNR by increasing the photon yield and decreasing photobleaching \cite{Kaufmann_2014}. However, fully utilizing the advantage of lower photobleaching kinetics requires longer illumination times, during which sample drift might become problematic. 

Experimentally, sample drift poses a challenge when recording SMLM data in all configurations, both at room and cryogenic temperatures. The drift distorts the recorded data and impacts the resolution of SMLM. In the context of a multi-parameter fit, any degradation or distortion of the measured intensity pattern might introduce errors in the localization \cite{deschout2012}. In addition, drift introduces an additional bias to the obtained dye positions in each frame, thus distorting the reconstructed SMLM localization map.
To retain superresolution on the nanometer scale, it is necessary to account for the drift both in the fitting procedure and post-processing of the data.

Sample drift can be caused by a variety of reasons, such as thermal fluctuations, vibrations and mechanical instabilities of the setup.
While sample motion during image acquisition should be reduced as much as possible, in practice it cannot be prevented completely, especially during the large acquisition times required for (cryo-)SMLM. One proposed method to compensate for the drift during acquisition is by means of closed feedback loops \cite{Coelho_2020, pertsinidis2010}. While this approach is common for correcting axial drift in order to keep the microscope in focus, implementations for lateral drift are rare. Real-time lateral drift compensation requires a dedicated piezoelectric sample stage that requires complex modifications to some microscopy setups. In the case of SMLM at cryogenic temperatures, the sample needs to be mounted on a cryostat, either in direct contact to liquid nitrogen or in a vacuum isolated environment, which, although possible \cite{liu2015}, considerably complicates the implementation of an active drift compensation stage.   

Most post-processing drift compensation methods can be grouped into methods that exploit the tracking of fiducial markers \cite{Lee_2012, Ma_2017, balinovic2019} and cross-correlation methods \cite{mlodzianoski2011,wang2014}. 
One advantage of cross-correlation methods is that they can be applied without the need of fiducial markers or any other modifications of the sample. However, they rely on the presence of features that remain visible over consecutive frames, which can be complicated due to blinking of the fluorophores. 
Fiducial tracking compensates the distortion of the SMLM localization map due to drift by tracking the sample movement during the experiment via fixed fiducial markers. The drift can then be corrected for by subtracting the measured fiducial track from the obtained localization coordinates \cite{Lee_2012, Ma_2017}. The signal from the fiducial markers is typically recorded at the same frame rate as the signal from the fluorophore emitters. In cryo-SMLM, however, this could be insufficient, as the low frame rate of down to 1Hz \cite{dahlberg2018} might lead to significant drifts during the illumination of one frame. 

\begin{figure}[!ht]
	\centering
	\includegraphics[width=\textwidth]{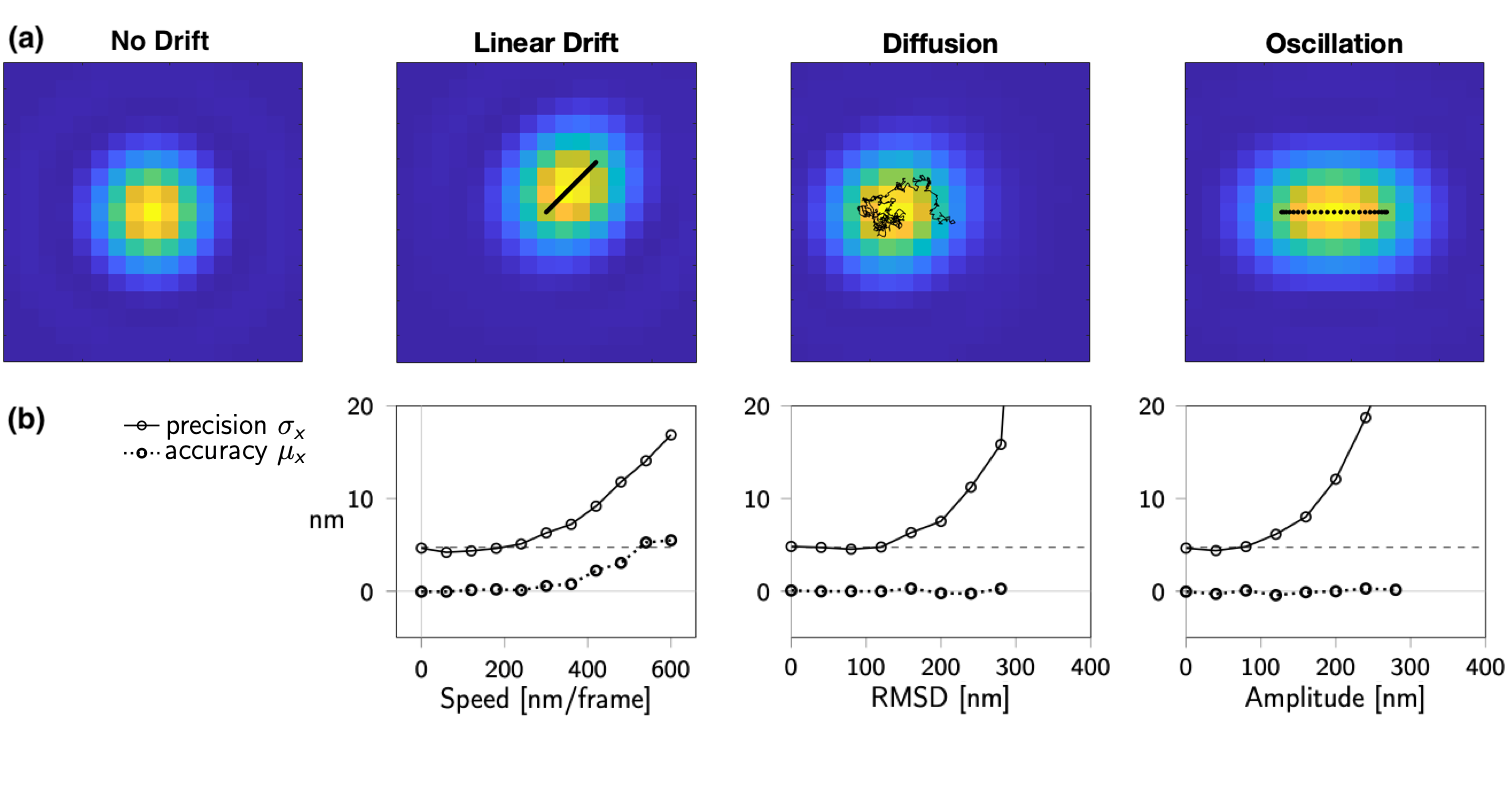}
	\caption{\textbf{Influence of drift on localization precision (Static fit).}
		\textbf{(a)} PSF images in case of no sample drift, linear drift with a speed of $\SI{400}{\nano\meter}/$frame, diffusion with a root-mean-square deviation (RMSD) of $\SI{300}{\nano\meter}/$frame and oscillation with an amplitude of $\SI{300}{\nano\meter}$ (left to right).
		\textbf{(b)} Localization precision (solid line) and bias (dotted line) for a static fitting procedure that only assumes knowledge about the centroid of the drift motion. The dashed horizontal line indicates the localization precision in case of no sample drift. As localization algorithm we use maximum-likelihood estimation (MLE) with the PSF model described in Section~\ref{sect_model}. Here, we use a sampling rate of the drift trajectory of $1$, corresponding to the centroid of the drift motion being known. We further assume here that this position is known exactly. For each data point we simulated fluorophores with random but fixed dipole orientations. Each data point represents the result of $1000$ simulations.}
	\label{fig_standardFit}
\end{figure}

If drift becomes very large, standard fiducial or cross-correlation correction methods reach their limitations. The measured intensity pattern then may deviate substantially from that of a static emitter. Localization procedures using a maximum-likelihood estimator of the position which assume a point-spread function (PSF) model without drift can then become unstable, in particular for cryo-SMLM. In Fig.~\ref{fig_standardFit} we demonstrate this effect. The first row shows the simulated intensity pattern of a fixed dipole emitter undergoing various forms of assumed sample drift, including linear drift, diffusion and oscillation. In the second row we show the computed localization errors for varying magnitudes of the sample drift: The precision $\sigma$ describes the spread of the errors whereas the accuracy $\mu$ is defined as the mean of the errors (see Methods for further details).
In case of drifts below \SI{100}{\nm}, the standard localization method without accounting for the drift trajectory performs remarkably well. However, in the case of large (nonlinear) drifts, the localization deteriorates very fast and the errors become very large. For example, the localization error exceeds \SI{20}{\nm} for an oscillation amplitude of \SI{300}{\nm}.

In this paper, we present a localization method for cryo-SMLM that accounts for distortions of the PSF due to sample drift directly in the fitting procedure. This approaches makes nanometer precision feasible even in case of very large sample drifts up to several hundred nanometers per frame.

\clearpage

\section{Results}\label{sect_results}

\begin{figure}[!ht]
	\centering
	\includegraphics[width=0.5\textwidth]{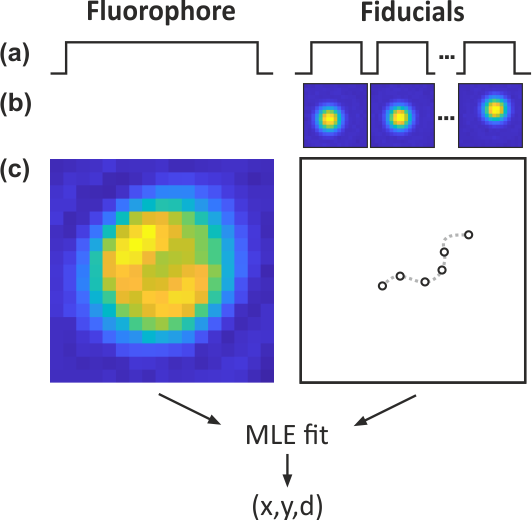}
	\vspace{10pt}
	\caption{\textbf{Fitting procedure accounting for drift.}
		The figure illustrates an overview of the experimental protocol assumed in the simulations, and the fitting procedure taking into account the drift trajectory of the sample.
		During the recording of one frame of the sample, several subframes of fiducial markers are recorded. Panel \textbf{(a)} shows the illumination protocols for the imaged fluorophore (left) and the fiducial markers (right). The resulting PSF images are shown in panel \textbf{(b)} for the fiducials, and panel \textbf{(c)} for the fluorophore. Localization of the fiducial markers yields an estimate of the drift trajectory (panel \textbf{(c)}, right). The gray line indicates the underlying ground truth drift trajectory, the black circles show the estimated position of the fiducial marker at the sampled time points. The estimated drift trajectory is incorporated in the MLE fit of the sample PSF, yielding an estimate of lateral position $(x,y)$ and defocus $d$. 
	}
	\label{fig_fitMethod}
\end{figure}

We use a maximum-likelihood estimator for 2D-position and defocus as described in Methods. In order to account for sample drift in the fitting procedure, an estimate of the drift trajectory during the image acquisition time is required. Fig.~\ref{fig_fitMethod} illustrates the illumination protocol and the image acquisition and fitting procedure. While the sample of interest is excited and imaged during one frame, several subframes are recorded of the fiducial markers. Thus, the position of the fiducial markers can be tracked on a smaller timescale, and an estimate of the drift trajectory within the SMLM frame is obtained. This drift estimate is then incorporated into the maximum-likelihood fit of the sample PSF.
If not mentioned otherwise, we assume in the following that we sample the drift at $25$ uniformly spaced time steps during one frame. We refer to this sampling rate as the trajectory sampling rate. In the following simulations, we further assume that the position of fiducial markers at these time steps is known exactly. Later, we will incorporate localization errors to capture a more realistic scenario.

\vspace{5pt}

In the following figures, we show results for the localization precision $\sigma_x$ and accuracy $\mu_x$ of the $x-$component for various simulation settings. The results for the $y$-component are similar and were omitted for clarity.

\begin{figure}[!ht]
	\centering
	\includegraphics[width=\textwidth]{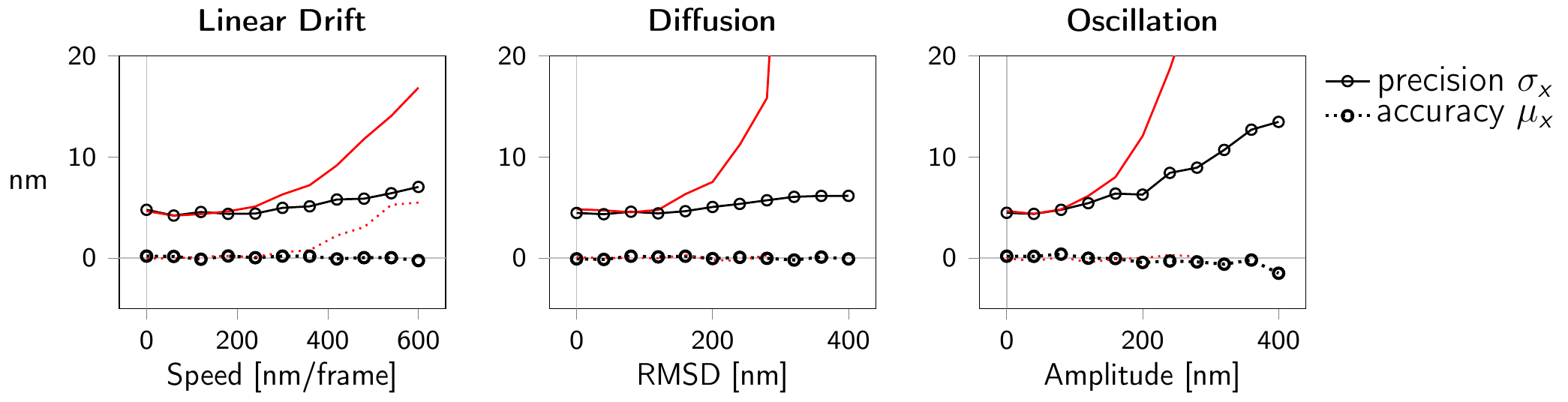}
	\caption{\textbf{Influence of drift on localization precision (dynamic and static fit).}
		Localization precision (solid line) and bias (dotted line) for the fit taking into account the drift trajectory. Left to right: linear drift, diffusion and oscillation. The red lines show the results obtained for the static fit (see Fig.~\ref{fig_standardFit}) for comparison.
		Each data point represents $500$ simulations.}
	\label{fig_compareMotions}
\end{figure}

In Fig.~\ref{fig_compareMotions} we compare the performance of our proposed fitting procedure including the drift trajectory against a fit that assumes only the centroid of the drift motion to be known (see Fig.~\ref{fig_standardFit}). For convenience, we refer to these two types of fit as \textit{dynamic fit} and \textit{static fit} in the following.
Again, we show the results for fixed dipole emitters undergoing three different types of sample movement, either linear drift, diffusion, or oscillation.
In order to quantify the performance, we calculate the localization precision $\sigma$ and accuracy $\mu$ of the fit.
In case of small sample movements, the dynamic and static fit show similar results. However, for large drifts, the dynamic fit yields superior fit results, showing stable localization results up to \SI{600}{\nm} for the linear drift, and up to \SI{400}{\nm} RMSD for diffusion. Only in the case of oscillation, the performance decreases slightly for large amplitudes, but is still highly superior to the static fit.

\begin{figure}[!ht]
	\centering
	\includegraphics[width=\textwidth]{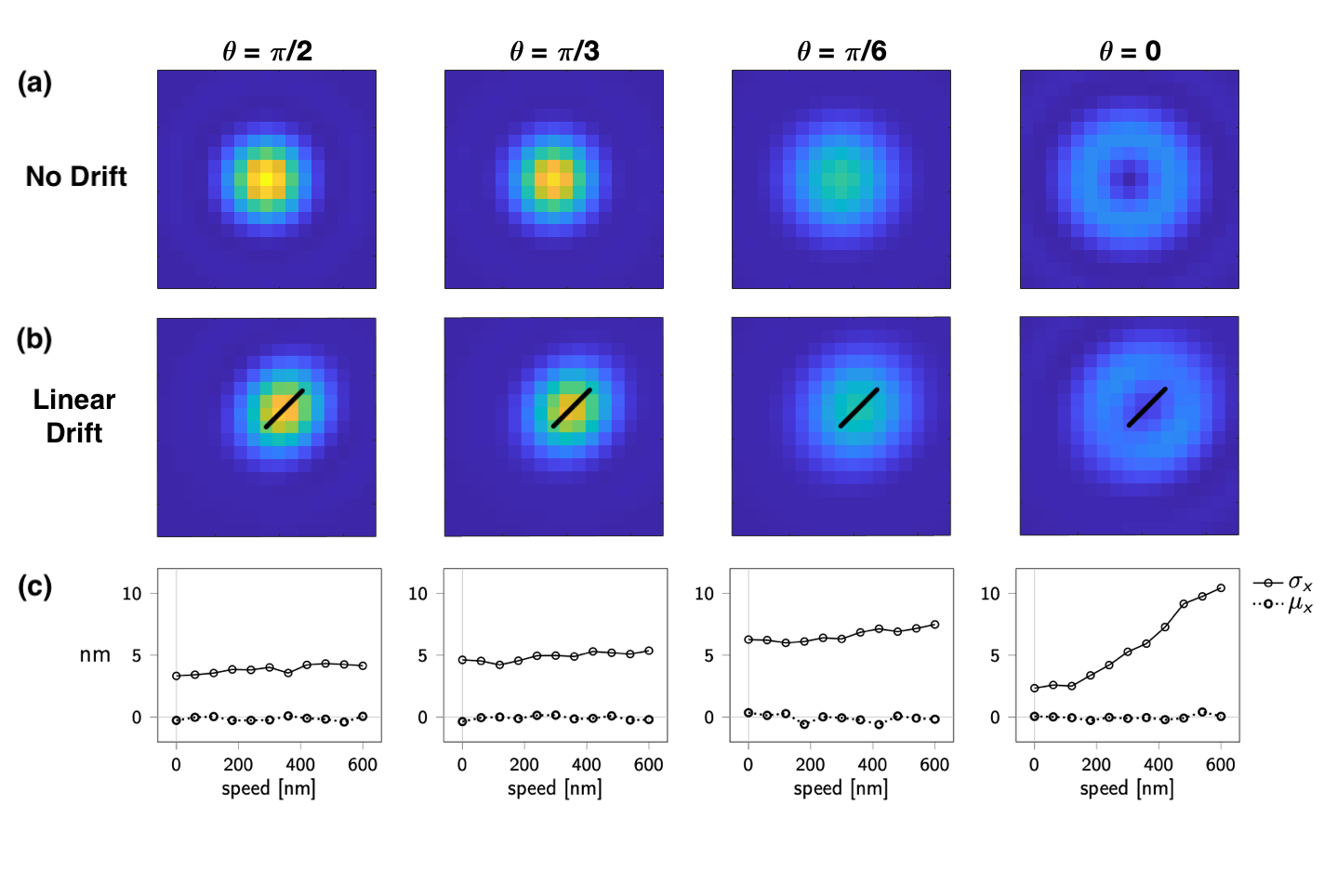}
	\caption{\textbf{Influence of inclination angle.}
		Simulations were carried out assuming a linear drift of $0-\SI{600}{\nano\meter}$ and fixed inclination angles $\theta=\pi/2, \pi/3, \pi/6, 0$ (left to right).
		Row \textbf{(a)} shows intensity distributions of emitters not undergoing any drift. Row \textbf{(b)} shows the intensity distributions simulated from emitters undergoing linear drift with a speed of $\SI{400}{\nano\meter}$ per frame. The drift trajectory is indicated by a black line. Row \textbf{(c)} shows the localization precision (solid line) and bias (dotted line) when performing a dynamic fit with $25$ trajectory sampling points. We assume a constant photon count of $10^5$ that is independent of orientation.  
		Each data point represents $500$ simulations.}
	\label{fig_compareInclinationAngles}
\end{figure}

Next, we examined the effect of dipole orientation on the performance of the localization procedure. In cryogenic conditions, the dipole of each emitter is fixed and the anisotropic emission pattern results in characteristic intensity patterns for each dipole orientation that may respond differently to sample drift. Fig.~\ref{fig_compareInclinationAngles} confirms this suspicion. The four panels show different assumed inclination angles of the fluorophore dipole ranging from $\theta=\pi/2$ (i.e., perpendicular to the optical axis) to $\theta=0$ (i.e.,  parallel to the optical axis). We assume that the emitter undergoes linear drift along the image diagonal, as indicated by the black lines in panel (b). The doughnut-shaped intensity pattern associated with dipoles parallel to the optical axis ($\theta=0$) is found to be more sensitive to large sample drifts.

\begin{figure}[!ht]
	\centering
	\includegraphics[width=0.4\textwidth]{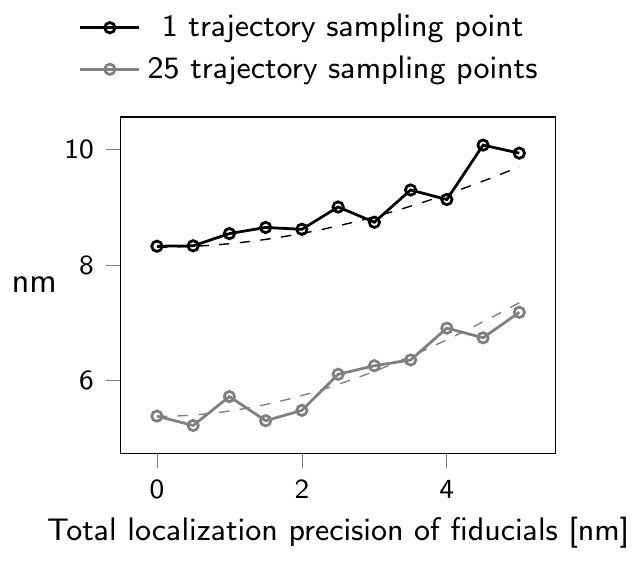}
	\caption{\textbf{Influence of fiducial localization precision.}
		Localization precision $\sigma_x$ for samples undergoing a linear drift (with a speed of $\SI{400}{\nm}$/frame) during the image recording process, where only a noisy estimate of the motion is known. We show results for different values of the total fiducial localization precision between $0$ (noise-free) and \SI{5}{\nm}, defined as the precision achieved with a trajectory sampling rate of $1$. For a sampling rate of $25$, the precision is adjusted accordingly, due to the overall constant photon budget being split up over $25$ frames. Results shown without subsampling of the trajectory (black line) and with $25$ trajectory sampling points (gray line). The dashed lines show the predicted precision. Each data point represents $1000$ simulations.}
	\label{fig_noisyMotion}
\end{figure}

So far, we assumed that we have an error-free estimate of the sample drift. In reality, the positions of the fiducial markers can only be estimated with a certain nonzero precision. In Fig.~\ref{fig_noisyMotion} we investigated quantitatively how these errors in the drift trajectory deteriorate the localization precision of the sample fluorophores.
For this, we vary the localization precision for the fiducial markers, $\sigma_{\text{fid}}$ between $0$ and \SI{5}{\nano\meter}. First, we simulate trajectories recorded at a sampling rate of $1$. As expected, the localization precision for the fluorophore, $\sigma_x$, scales with $\sigma_{\text{fid}}$ according to $\sigma_x = \sqrt{\sigma_0^2 + \sigma_{\text{fid}}^2}$\,, where  $\sigma_0$ is the localization precision for the fluorophore in the absence of trajectory noise. 
Next, we are interested whether a higher trajectory sampling rate improves the results. In practice, this implies a deteriorated localization precision $\sigma_{\text{fid}}$, since the overall photon budget for each fiducial marker is divided into several subframes. For example, a trajectory sampling rate of $25$ results in a $\sigma_{\text{fid}}$ that is deteriorated by a factor of $5$ (cf.\ equation (5) in \cite{thompson2002}). To facilitate the comparability of the two sampling rates, on the x-axis we plot the total localization precision, which would be achieved if the total photon budget of the fiducial marker was recorded in a single frame. Interestingly, even in case of a 5-fold deteriorated localization precision we observe an improved localization precision for the single fluorophore, most likely since the full drift trajectory could now be considered in the analysis. 

\begin{figure}[!ht]
	\centering
	\includegraphics[width=0.4\textwidth]{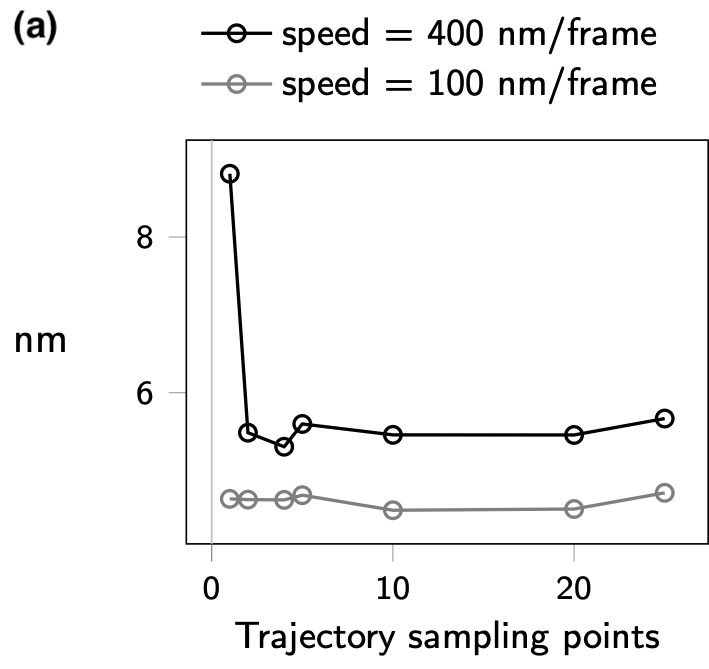}  \includegraphics[width=0.4\textwidth]{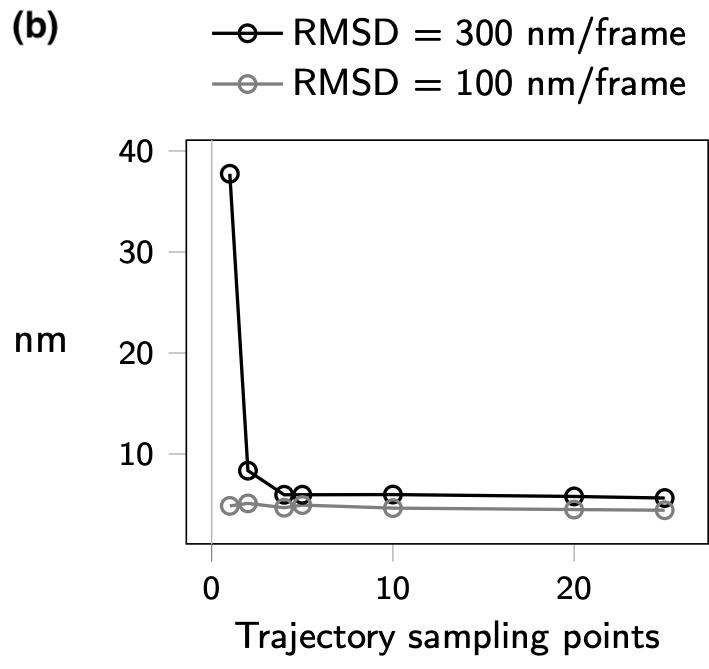}
	\caption{\textbf{Influence of the number of trajectory sampling points.}
		The number of trajectory sampling points is varied from $1$ to $25$.
		\textbf{(a)} linear drift with a speed of $\SI{100}{\nm}$ and $\SI{400}{\nm}$ per frame. \textbf{(b)} diffusion with a RMSD of $\SI{100}{\nm}$ and $\SI{300}{\nano\meter}$ per frame. 
		The total localization precision of fiducials is set to $\SI{1}{\nm}$ and is adjusted w.r.t. the number of trajectory sampling points.
		Each data point represents $2500$ simulations.}
	\label{fig_fiducial_framerate}
\end{figure}

Up to now, we only considered a trajectory sampling rate of either $1$ or $25$. In Fig.~\ref{fig_fiducial_framerate}, we investigate the influence of a range of different sampling rates. In panel (a), we assume linear drift with a speed of $\SI{100}{\nano\meter}$ and $\SI{400}{\nano\meter}$ per frame. At a drift speed of $\SI{100}{\nano\meter}$ per frame, no gain in localization precision is achieved by increasing the sampling rate. However, in the case of the larger drift of $\SI{400}{\nano\meter}$ per frame, the localization precision improves significantly when increasing the sampling rate to $2$. Surprisingly, a further increase of the sampling rate was not found to yield any further improvements. In panel (b), we assume diffusion with an RMSD of $\SI{100}{\nano\meter}$ and $\SI{300}{\nm}$ per frame. Similarly, a sampling rate of $1$ causes large errors in the case of the larger drift. Here, a sampling rate of $4$ or greater is producing identical results. 

\begin{figure}[!ht]
	\centering
	\includegraphics[width=0.48\textwidth]{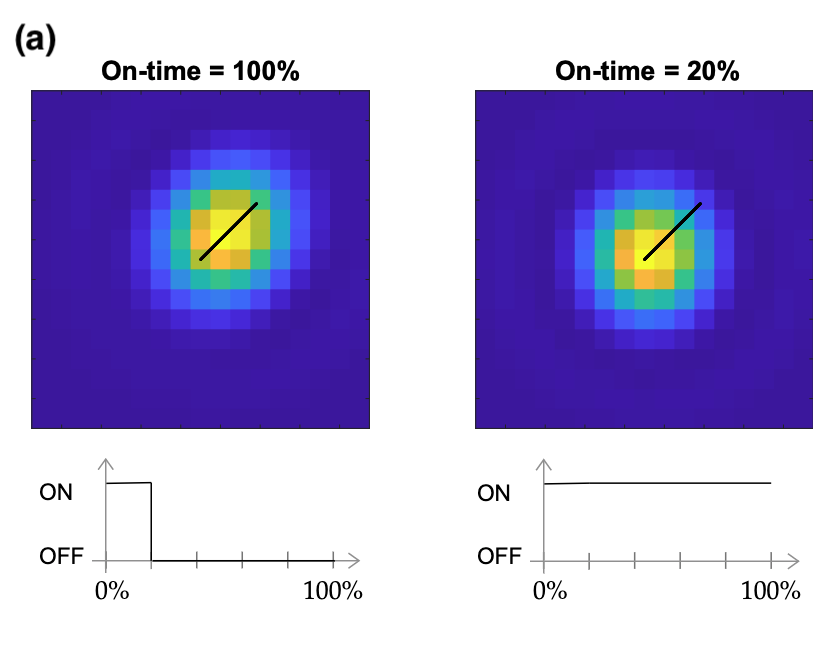}
	\includegraphics[width=0.48\textwidth]{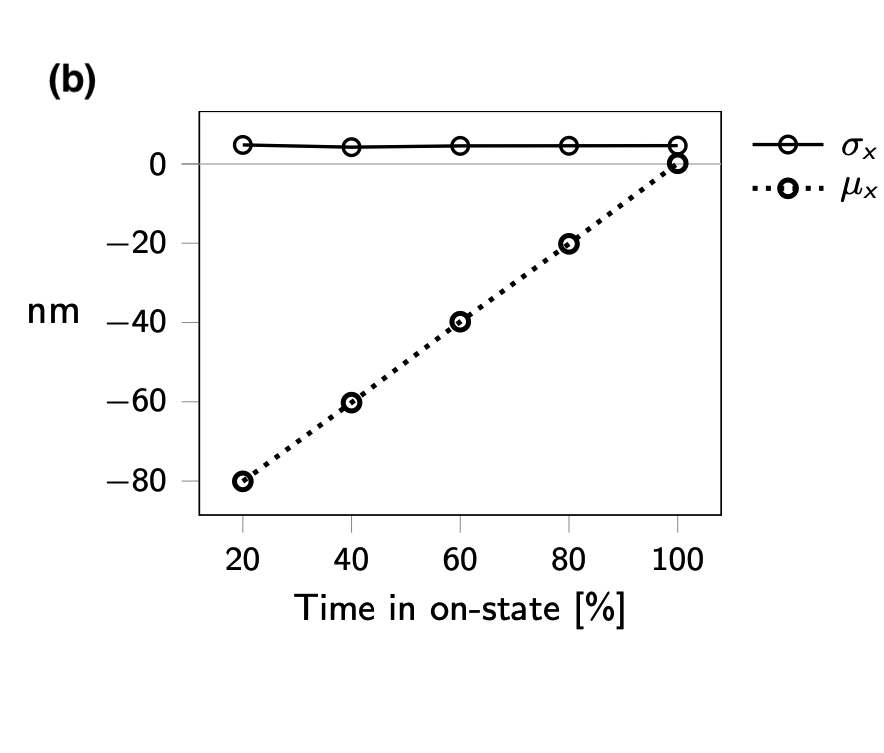}
	\caption{\textbf{Influence of off-switching of fluorophore during frame.}
		We assume the fluorophore to be active at the beginning of the illumination, and to switch to the dark state after $20$ to $100\%$ of the illumination time. \textbf{(a)} Intensity patterns of two emitters undergoing identical linear drift. The emitter in the left image is continuously on, while the emitter in the right image switches to the off-state after $20\%$ of the frame recording time. \textbf{(b)} Arising localization precision $\sigma_x$ and accuracy $\mu_x$ for a drift of $\SI{200}{\nm}$ per frame. 
		Each data point represents $1000$ simulations.}
	\label{fig_offSwitching}
\end{figure}

Finally, in Fig.~\ref{fig_offSwitching}, we examine the scenario where an emitter switches off during the image acquisition of a single frame. For the drift scenario, we choose linear drift with a speed of $\SI{200}{\nano\meter}$ per frame. We assume the emitter to be active at the beginning of the recording of the frame, and to switch off at a time point varying from $20\%$ to $100\%$ of the total recording time for the single frame. As expected, this transition introduces a very large localization bias.

\section{Discussion}\label{sect_conclusion}

In this paper, we investigated the impact of sample drift on the determination of fixed dipole positions from SMLM data. We considered an astigmatic PSF model for localization, and simulation parameters typical for cryo-SMLM data.  

In typical experimental setups, drift cannot be avoided completely. Large drifts during acquisition times can occur during cryo-SMLM measurements, as they tend to require higher exposure times and might have additional drift caused by thermal gradients or liquid nitrogen bubbling \cite{dahlberg2018}. In these cases, traditional post-processing drift correction approaches might not be sufficient to retain resolution at the nanoscale. If no active drift correction during image acquisition is applied \cite{Coelho_2020, pertsinidis2010}, the drift should be corrected in post-processing of the data, either based on fiducial tracking or cross-correlation \cite{Lee_2012, Ma_2017, balinovic2019, mlodzianoski2011, wang2014}. While cross-correlation cannot be applied to correct for drift occurring within a frame, fiducial markers can be recorded with a higher frame rate, thus yielding information of the drift trajectory on a sub-frame timescale.
The commonly used strategy of imaging the drift trajectory with the same frame rate as the sample was found to be sufficient to achieve nanometer precision in the case of low to moderate sample drift. This method was found to be surprisingly stable even with drift magnitudes of up to $\SI{200}{\nano\meter}$.  However, for very large drifts, a significant degradation of the localization precision was detected, indicating the necessity of more sophisticated correction.  

Along this line, a conceptually similar approach to the one explored here has been reported in \cite{dahlberg2018}. There, the drift correction is preceded by a binning step to improve the SNR. A key difference is the way in which the measured drift enters the method and the choice of the PSF model. In reference \cite{dahlberg2018}, an elliptical Gaussian PSF model was assumed and the drift was subtracted from the data before fitting. In contrast, here we use a full vectorial PSF model and leave the data unaltered, while incorporating the measured drift into the PSF model for fitting. 

For our method, we sample the drift trajectory at a higher rate, which was found to restore a localization precision that is similar to what could be achieved without drift, retaining nanometer precision even in the presence of drift of several hundred nanometers. This was showcased for three different types of drift that are experimentally most relevant: linear drift, diffusion and oscillation. For the rather large magnitudes of drift considered here, a trajectory sampling rate of $4$ was determined to be sufficient.

Note that the photostability of the fluorophore within a frame is important in order for our method to yield accurate results. Off-switching of an emitter within the acquisition of a single frame was found to induce a substantial localization bias. In an experiment, however, the typical on-time of a fluorophore exceeds the duration of a single frame, and the fluorophore will be imaged throughout multiple consecutive frames. Hence, this issue can be avoided by discarding the first and last frame, in which the fluorophore was detected, from the analysis. Alternatively, the time point of on- or off-switching can be estimated from the intensity in the affected frame compared to the intensity of the on-frames.   

In particular for large drifts, the analysis region for the fitting needs to be chosen large enough such that it contains the entire signal. This requires sufficient spatial separation of individual emitter signals and, in particular, no signal overlap.

In summary, we have demonstrated that the standard fiducial correction of sample drift achieves good results for low to moderate levels of sample drift. In case of very large drifts, increasing the sample rate of the fiducials and accounting for the drift trajectory in the fitting procedure restores nanometer precision, allowing for very long image acquisition times for individual frames in cryo-SMLM.

\section{Methods}

\subsection{Mathematical Model}\label{sect_model}

We use the same image formation model as in our previous publication \cite{HinSch_2022}, which in turn is based on \cite{Axelrod_2012}. A dipole point source can be characterized by its orientation $(\theta, \phi)$ and its position $(x, y,z)$, where $\theta$ denotes the inclination angle with respect to the optical axis, and $\phi$ denotes the azimuthal angle within the sample plane w.r.t to an arbitrary but fixed coordinate system. 
For our model, we consider a fluorophore positioned in the focal plane ($z = 0)$ with a fixed dipole orientation. In addition, we consider a possible unknown defocus $d$ of the objective.

The dipole emission pattern can be expressed as an angular spectrum of plane waves. These plane waves, while propagating through the optical system, are refracted and reflected according to Snell's law and Fresnel equations. The infinity-corrected objective captures the emission light emanating radially from the source and directs it parallel to the optical axis through the back focal plane of the objective. We consider the electric field in the back focal plane (BFP) in Cartesian coordinates:
	\begin{equation}
		\fieldBPF(x_b,y_b) = \fieldBPF(x_b,y_b; \theta, \phi).
	\end{equation}
In this model, the BFP field only depends on the orientation of the emitter, as the lateral position will be modeled via tip-/tilt aberrations. Any wavefront deformation, either caused by aberration or deliberate distortion, is modeled by introducing additional phase factors \cite{Goodman}. We therefore define the aberration term $ W_\xi(x,y) $ modeling tip/tilt aberrations and defocus (and possibly additional aberrations). The parameter vector $\xi =(x, y, d)$ collects the position of the emitter and the defocus, the parameters which we are later interested in estimating. 

The light beam then enters the tube lens as an infinite parallel beam. The field at the focal plane of the tube lens is given by the Fourier transform of $\fieldBPF$ multiplied with the phase factor introduced above,
	\begin{equation} \label{psf_field}
		E_\xi (x_f, y_f) = \frac{1}{i\lambda f}e^{\frac{ik}{\lambda f}(x_f^2+y_f^2)}\int \fieldBPF(x_b,y_b) e^{\frac{2 \pi i}{\lambda} W_\xi(x_b,y_b)} e^{-\frac{2 \pi i}{\lambda f}(x_b x_f + y_b y_f)} dx_b \, dy_b \, , 
	\end{equation}
where $f$ denotes the focal length of the tube lens and the subscript $f$ indicates coordinates in the focal plane.
The intensity distribution in the focal plane of the tube lens is given by the absolute value of the electric field,
	\begin{equation} \label{psf_intensity}
		I_\xi(x_f,y_f) = |E_\xi(x_f,y_f)|^2.
	\end{equation}
We now consider the scenario where the emitter undergoes a lateral motion during the recording process. To model this motion, we define the path
	\begin{equation}
		\gamma: [0,T] \to \Omega \subset \R^2.
	\end{equation}
The intensity measured at the detector plane is then the integrated intensity 
	\begin{equation} \label{intensity_motion}
		\I_{\xi,\gamma} (x_f, y_f):= \int_0^T  |E_{\xi, \gamma(t)} (x_f,y_f)|^2 dt, 
	\end{equation}
where
	\begin{equation} \label{psf_gamma}
		E_{\xi, \gamma(t)} (x_f,y_f) := \frac{1}{i\lambda f}e^{\frac{ik}{\lambda f}(x_f^2+y_f^2)}  \int E_{\text{BFP}}(x_b, y_b) e^{ \frac{2 \pi i}{\lambda} W_{\xi,\gamma(t)} (x_b, y_b)} e^{-\frac{2 \pi i}{\lambda f}(x_b x_f + y_b y_f)} d x_b \, d y_b . 
	\end{equation}
Here, the motion $\gamma(t)$ at time $t$ is incorporated into the aberration term $W_{\xi,\gamma(t)}$ as additional tip and tilt terms, 
	\begin{equation} \label{wavefront}
		W_{\xi, \gamma(t)}(x_b,y_b) = W_\xi(x_b,y_b) + \gamma_1(t) Z_2(x_b,y_b) + \gamma_2(t) Z_3(x_b,y_b),
	\end{equation}
where $Z_2$ and $Z_3$ denote the second and third Zernike polynomial in Noll's indices \cite{Noll_1976}. The remaining aberration term $W_\xi$ is further expanded into Zernike polynomials $Z_2, Z_3$ and $Z_4$, modelling the position of the fluorophore and the defocus, both specified by the parameter $\xi$. 

For simulation and fitting purposes, we consider \eqref{intensity_motion} using a discrete representation of $\gamma$.  We assume that we sample the motion at $N$ uniformly spaced time steps, resulting in a measurement $\hat{\gamma}\in \Omega^N$. We can approximate the integral in  \eqref{intensity_motion} by the sum 
	\begin{equation} \label{intensity_motion_discrete}
		\mathcal{I}_{\xi, \hat{\gamma}}(x_f, y_f) \equiv \sum_{k=1}^N \left| \int \fieldBPF(x_b,y_b) e^{\frac{2 \pi i}{\lambda} W_{\xi, \hat{\gamma}_k} (x_b,y_b)} e^{-\frac{2 \pi i}{\lambda f}\left(x_b x_f + y_b y_f \right)}dx_b \, dy_b  \right|^2 , 
	\end{equation}
which is the superposition of the signal of $N$ emitters and corresponds to \eqref{intensity_motion} being approximated with a midpoint rule at the supporting points $\left(\hat{\gamma}_k \right)_{k=1}^N$. The desired amplitude of \eqref{intensity_motion_discrete} will be introduced with an appropriate scaling factor.

\subsection{Simulations}

Unless otherwise specified, we use the same set of parameters and assumptions as previously described in \cite{HinSch_2022}, the key aspect of which is an astigmatic and low-aperture (NA=$0.7$) imaging model.

We model an air objective ($n_2=1$) with a magnification of 60x and a focal length of \SI{3}{\mm}, and a tube lens with a focal length of $f = \SI{180}{\nm}$. We assume a biological sample with a refractive index of water ($n_2=1.33$), and dyes with an emission wavelength of $\lambda=\SI{680}{\nm}$.

As input parameters for our simulated data we randomly sample values for the emitter's position $(x^*, y^*)$ and dipole orientation $(\theta^*, \phi^*)$ as well as defocus $d^*$ from uniform distributions. We use the superscript $^*$ to denote a ground truth variable. Position is sampled within $\pm 1$ pixel from the center (corresponding to the optical axis). Orientation is selected within the intervals $\theta^* \in [0,\pi]$ and $\phi^* \in [0,2\pi]$. The defocus is chosen randomly between $\pm \SI{500}{\nano\meter}$.

We consider three different types of sample drift: Diffusion, linear drift in direction of $\pi/4$ w.r.t.\ to the x-axis and oscillation in x-direction. For the oscillation, we use a frequency of $10$ oscillations per frame. 

We model the astigmatic distortion by adding an additional Zernike polynomial $Z_6$ to the phase factor \eqref{wavefront} and setting the corresponding Zernike coefficient to $0.11 \lambda$.

For the simulation of the data, we calculate the intensity distribution \eqref{intensity_motion_discrete} within an analysis region of $17 \times 17$ pixels using $N=500$ discretization points in time. To reduce numerical inaccuracies, we calculate the intensity distribution on a subpixel grid and then sum up over the individual pixel bins. We use a pixel size of $\SI{108}{\nano\meter}$ and an oversampling factor of $9$ for the subpixel grid. We scale the total intensity (i.e. the sum of all pixel values) to the desired photon count and then apply Poissonian shot noise. As photon count we select $10^5$ photons in all simulations. 

All simulations were carried out in MATLAB.

\subsection{Parameter estimation}

Given a possibly noisy input image $\I$, we want to retrieve an estimate of the parameter vector $\xi^* $, containing the position of the emitter and the defocusing of the optical system.  

First, we subtract the mean background signal $b^2$ from the image, which we estimate from the mean signal of an image containing no fluorophore signal.
The total number of detected photons from a fluorophore is estimated by summing over all pixels of the noise-corrected image.
For the following fitting procedure, we normalize the photon count in the images.
As additional input, we require an estimate $\hat{\gamma}$ of the motion and an estimate $(\hat{\theta}, \hat{\phi})$ of the dipole orientation of the emitter.
We assume that the errors for both the inclination and azimuthal angle are distributed normally with mean 0 and variance 2°.
The estimate $\hat{\gamma}$ of the motion is possibly contaminated with noise, which is assumed to follow a Gaussian distribution.
We proceed by minimizing the negative log-likelihood using the Matlab function $\emph{fminunc}$. The log-likelihood function is given by 
	\begin{equation} \label{loglikelihood}
		\ell(\xi) = \sum \left( \I \cdot \ln( \I_{\xi, \hat{\gamma}}) - \I_{\xi, \hat{\gamma}} - \ln(\I!) \right)
	\end{equation} 
where $ \I_{\xi, \hat{\gamma}}$ is the pixelated intensity distribution \eqref{intensity_motion_discrete} with the parameters $\xi$. Note that all expressions in \eqref{loglikelihood} are matrix-valued and the operations, including multiplication, should be understood component-wise. The sum is then performed over all matrix elements. We refer to Eq. $(27)$ in \cite{Ober_2004} for a derivation of the likelihood function. By applying the logarithm, we then arrive at the log-likelihood function.
For the fit, the normalized image $\I_\xi$ of the PSF is calculated using an oversampling factor of 3.
As initial guess for the optimization procedure we select  randomly chosen values from the admissible set of parameters. Minimizing the negative log-likelihood yields an estimate 
	\begin{equation}
		\hat{\xi} = ( \hat{x}, \hat{y}, \hat{d}) = \operatorname{argmin}\!\left( - \ell(\xi)\right).
	\end{equation}

We repeat the simulation and fitting procedure $n$ times for each data point. The quantities we are interested in are the localization accuracy, which is defined by the mean  error $\mu_x = \frac{1}{n}\sum (\hat{x}-x^*) \,$ and the localization precision $\sigma_x$, which is defined as the standard deviation of the estimates from the mean, $\sigma_x^2 := \frac{1}{n}\sum (\hat{x}-x^*-\mu_x)^2 $. The quantities for the $y$-position and defocus are defined analogously. Note that the localization precision and accuracy are typically defined for an isolated emitter that is imaged $n$ times with only the realization of the shot noise differing in each frame \cite{Deschout_2014}. In contrast, we calculate the precision and accuracy for $n$ emitters, where the underlying parameters are randomly sampled. Since we are mainly interested in an overall precision, we believe this quantity to be more relevant for our purposes.

\section*{Acknowledgments}
F. Hinterer, S. Hubmer, and R. Ramlau were funded by the Austrian Science Fund (FWF): F6805-N36. M.C. Schneider, M. Lopez-Martínez, and G.J. Schütz were funded by  the Austrian Science Fund (FWF): F6809-N36.

\section*{Data availability statement}
The data that support the findings of this study are available from the corresponding author upon reasonable request.

{\footnotesize
\bibliographystyle{plain}
\bibliography{mybib}
}

\end{document}